\documentclass[]{mn2e}
\usepackage{times}
\usepackage{amsmath}
\usepackage{amssymb,amsfonts}
\usepackage{graphicx}
\usepackage{epstopdf}
\usepackage{verbatim}

\def\gapprox{\lower.4ex\hbox{$\;\buildrel >\over{\scriptstyle\sim}\;$}}
\def\lapprox{\lower.4ex\hbox{$\;\buildrel <\over{\scriptstyle\sim}\;$}}
\def\be{\begin{equation}}
\def\be{\begin{equation}}
\def\ee{\end{equation}}
\def\bea{\begin{eqnarray}}
\def\eea{\end{eqnarray}}
\catcode`\@=11 \font\tenmib=cmmib10 \font\tensyb=cmbsy10
\font\tenbi=cmmib10
\newfam\bifam 
\textfont\bifam=\tenbi  

\def\unboldmath{\everymath{}\everydisplay{}
          \textfont\@ne\teni
          \textfont\tw@\tensy
          }
\def\boldmath{$\!\!$\relax\everymath{\mit}\everydisplay{\mit}
        \textfont\@ne\tenmib
        \textfont\tw@\tensyb
        \relax}%
\catcode`\@=12

\begin{document}
\tolerance=10000

\def\lesssim{\mathrel{\hbox{\rlap{\hbox{\lower4pt\hbox{$\sim$}}}\hbox{$<$}}}}
\def\gtrless{\mathrel{\hbox{\rlap{\hbox{\lower3pt\hbox{$<$}}}\hbox{$>$}}}}

\def\cenx4{{Cen~X$-$4}}
\def\aql{{Aql~X$-$1}}
\def\1e{{1E 1207.4-5209}}
\def\exo{{EXO 0748-676}}
\def\saxj{{SAX J1808.4$-$3658}}
\newcommand{\ud}[2]{\mbox{$^{+ #1}_{- #2}$}}

\def\be{\begin{equation}}
\def\ee{\end{equation}}
\def\bea{\begin{eqnarray}}
\def\eea{\end{eqnarray}}
\def\c{\cite}

\def\et{ {\it et al.}}
\def\lan{ \langle}
\def\ran{ \rangle}
\def\ov{ \over}
\def\ep{ \epsilon}

\def\et{ {\it et al.}}
\def\la{ \langle}
\def\ra{ \rangle}
\def\ov{ \over}
\def\ep{ \epsilon}

\def\mdot{\ifmmode \dot M \else $\dot M$\fi}    
\def\mxd{\ifmmode \dot {M}_{x} \else $\dot {M}_{x}$\fi}
\def\med{\ifmmode \dot {M}_{Edd} \else $\dot {M}_{Edd}$\fi}
\def\bff{\ifmmode B_{f} \else $B_{f}$\fi}

\def\apj{\ifmmode ApJ\else ApJ\fi}    
\def\apjl{\ifmmode  ApJ\else ApJ\fi}    %
\def\aap{\ifmmode A\&A\else A\&A\fi}    %
\def\mnras{\ifmmode MNRAS\else MNRAS\fi}    %
\def\nat{\ifmmode Nature\else Nature\fi}
\def\prl{\ifmmode Phys. Rev. Lett. \else Phys. Rev. Lett.\fi}
\def\prd{\ifmmode Phys. Rev. D. \else Phys. Rev. D.\fi}

\def\ms{\ifmmode {\rm M_{\odot}} \else ${\rm M_{\odot}}$\fi}    
\def\na{\ifmmode \nu_{A} \else $\nu_{A}$\fi}    
\def\nk{\ifmmode \nu_{K} \else $\nu_{K}$\fi}    
\def\ns{\ifmmode \nu_{{\rm s}} \else $\nu_{{\rm s}}$\fi}
\def\no{\ifmmode \nu_{1} \else $\nu_{1}$\fi}    
\def\nt{\ifmmode \nu_{2} \else $\nu_{2}$\fi}    
\def\ntk{\ifmmode \nu_{2k} \else $\nu_{2k}$\fi}    
\def\dnmax{\ifmmode \Delta \nu_{max} \else $\Delta \nu_{2max}$\fi}
\def\ntmax{\ifmmode \nu_{2max} \else $\nu_{2max}$\fi}    
\def\nomax{\ifmmode \nu_{1max} \else $\nu_{1max}$\fi}    
\def\nn{\ifmmode \nu_{\rm NBO} \else $\nu_{\rm NBO}$\fi}    
\def\nh{\ifmmode \nu_{\rm HBO} \else $\nu_{\rm HBO}$\fi}    
\def\nqpo{\ifmmode \nu_{QPO} \else $\nu_{QPO}$\fi}    
\def\nz{\ifmmode \nu_{o} \else $\nu_{o}$\fi}    
\def\nht{\ifmmode \nu_{H2} \else $\nu_{H2}$\fi}    
\def\ns{\ifmmode \nu_{s} \else $\nu_{s}$\fi}    
\def\nb{\ifmmode \nu_{{\rm burst}} \else $\nu_{{\rm burst}}$\fi}
\def\nkm{\ifmmode \nu_{km} \else $\nu_{km}$\fi}    
\def\ka{\ifmmode \kappa \else \kappa\fi}    
\def\dn{\ifmmode \Delta\nu \else \Delta\nu\fi}    

\def\vk{\ifmmode v_{k} \else $v_{k}$\fi}    
\def\va{\ifmmode v_{A} \else $v_{A}$\fi}    %
\def\vf{\ifmmode v_{ff} \else $v_{ff}$\fi}    

\def\rs{\ifmmode {R_{s}} \else $R_{s}$\fi}    
\def\ra{\ifmmode R_{A} \else $R_{A}$\fi}    
\def\rso{\ifmmode R_{S1} \else $R_{S1}$\fi}    
\def\rst{\ifmmode R_{S2} \else $R_{S2}$\fi}    
\def\rmm{\ifmmode R_{M} \else $R_{M}$\fi}    
\def\rco{\ifmmode R_{co} \else $R_{co}$\fi}    
\def\ris{\ifmmode {R}_{{\rm ISCO}} \else $ {\rm R}_{{\rm ISCO}} $\fi}
\def\rsix{\ifmmode {R_{6}} \else $R_{6}$\fi}
\def\rinfty{\ifmmode {R_{\infty}} \else $R_{\infty}$\fi}
\def\rinfsix{\ifmmode {R_{\infty6}} \else $R_{\infty6}$\fi}

\def\rxj{\ifmmode {RX J1856.5-3754} \else RX J1856.5-3754\fi}
\def\1739{\ifmmode {XTE  J1739-285} \else XTE  J1739-285\fi}

\title{Kernel Regression For Determining Photometric Redshifts From Sloan Broadband Photometry }

\author[D. Wang et al.]{D. Wang$^{1,2}$, Y. X. Zhang$^{1}$, C. Liu$^{1,2}$ Y. H.
Zhao$^{1}$ \\
1. National Astronomical Observatories,
 Chinese Academy of Sciences, Beijing 100012,
dwang@lamost.org \\2. Graduate University of Chinese Academy of
Sciences, Beijing 100080, China}

\date{\today}

 \maketitle

 \begin{abstract}
 We present a new approach, kernel regression, to
 determine photometric redshifts for 399,929 galaxies in the
 Fifth Data Release of the Sloan Digital Sky Survey (SDSS). In our case, kernel
 regression is a weighted average of spectral redshifts of the neighbors for a query point,
 where higher weights are associated with points that are closer to the query point. One
important design decision when using kernel regression is the choice
of the bandwidth. We apply 10-fold cross-validation to choose the
optimal bandwidth, which is obtained as the cross-validation error
approaches the minimum. The experiments show that the optimal
bandwidth is different for diverse input patterns, the least rms
error of photometric redshift estimation arrives at 0.019 using
color+eClass as the inputs, the less rms error amounts to 0.020
using $ugriz$+eClass as the inputs. Here eClass is a galaxy spectra
type. Then the little rms scatter is 0.021 with color$+r$ as the
inputs. As a result, except the parameters (e.g. magnitudes and
colors), eClass is a valid parameter to predict photometric
redshifts. Moreover the results also suggest that the accuracy of
estimating photometric redshifts is improved when the sample is
divided into early-type galaxies and late-type ones, especially for
early-type ones, the rms scatter amounts to 0.016 with color+eClass
as the inputs. In addition, kernel regression achieves high accuracy
to predict the photometric eClass ($\sigma_{\rm rms}$ = 0.034) using
color$+r$ as the input pattern. For kernel regression, the more
parameters considered, the accuracy of photometric redshifts is not
always higher, but satisfactory only when appropriate parameters are
chosen. Kernel regression is comprehensible and accurate regression
models of the data. Experiments reveal the superiority of kernel
regression when compared to other empirical training approaches.

\end{abstract}


\section{Introduction}

In general, the redshifts of galaxy are measured spectroscopically.
In order to achieve high signal-to-noise spectra, long integration
time is required. For those large and faint sets of galaxies,
however, spectra of galaxies are not easy or impractical to obtain.
In the absence of spectroscopic data, redshifts of galaxies may be
estimated using medium- or broadband photometry, which may be
thought of as very low-resolution spectroscopy. Though such
photometric redshifts are necessarily less accurate than true
spectroscopic redshifts, they nonetheless are sufficient to
determine the formation and evolution properties of large number of
galaxies rather than to study accurate redshift of individual galaxy
(Gwyn 1990). Photometric redshifts may be obtained less expensively
and for much larger samples than is possible with spectroscopy. In
the nineties, photometric redshifts is rapidly becoming a crucial
tool in mainstream observational cosmology. To date, some
photometric redshift catalogs have been used to deal with several
scientific issues, e.g. the evolution of the luminosity density and
the number of massive galaxies already assembled at early epochs
(Fontana et al. 2000), the evolution of galaxy size (Poli et al.
1999; Giallongo et al. 2000), the determination of cosmological
baryonic and matter densities (Blake et~ al. 2007), and the
clustering of luminous red galaxies in SDSS imaging data
(Padmanabhan et al. 2007) .

Techniques for deriving photometric redshifts were pioneered by Baum
(1962). Subsequent implementations of these basic techniques have
been made by Couch et al. (1983) and Koo (1985). Photometric
redshift techniques have been divided into two broad categories:
template matching method and empirical training-set method. There
are advantages and disadvantages to each approach. The former
approach relies on fitting model galaxy spectral energy
distributions (SEDs) to the photometric data, where the models span
a range of expected galaxy redshifts and spectral types (e.g.,
Sawicki, Lin \& Yee 1997). A library of template spectra (e.g.
Bruzual \& Charlot 1993; Coleman, Wu \& Weedman 1980) are employed.
A $\chi^{2}$ fit is used to obtain the optimal template pairs for
each galaxy. The various techniques in this kind is different from
their choice of template SED's and in the procedure for fitting.
Template SED's may come from population synthesis models (eg.
Bruzual \& Charlot 1993) or from spectra of real objects (eg.
Coleman, Wu \& Weedman 1980). Both kinds of templates have their
weaknesses - template SED's from population synthesis models may
include unrealistic combinations of parameters or exclude known
cases. The real galaxy templates are almost always derived from data
on bright low redshift galaxies, and may be poor representations of
the high redshift galaxy population (Wadadekar 2005).

The latter approach depends on using an existing spectroscopic
redshift sample as a training set to derive photometric redshifts
as the function of photometric data. Some typical training-set
methods employed include: artificial neural networks (ANNs,
Colister \& Lahav 2004; Firth, Lahav \& Somerville 2003; Vanzella
et al. 2004; Li et al. 2006), support vector machines (SVMs,
Wadadekar 2005), ensemble learning and Gaussian process regression
(Way \& Srivastava 2006) and linear and non-linear polynomial
fitting (Brunner et al. 1997; Wang, Bahcall, \& Turner 1998;
Budav$\acute{a}$ri et al. 2005; Hsieh et al. 2005; Connolly et al.
1995). Such techniques have strengthes that they are automatically
constructed by the properties of galaxies in the real universe and
require no additional assumptions about their formation and
evolution. However for the empirical best fit method, such as
linear and non-linear polynomial fitting, it is difficult to
extrapolate to objects fainter than the spectroscopic limit. For
the ANN approach, its optimal architecture is not easy to obtain,
moreover and it is easy to get stuck in local minima during
training stage. Unlike ANNs, SVMs do not need choice of
architecture before training, but the optimal parameters in their
models are obtained with much effort.

Another interpolative training-set methods are instance-based
learning techniques, applied to predict photometric redshifts (eg.
Csabai et al. 2003; Ball et al. 2007). Instance-based learning
methods base their predictions directly on (training) data that
has been stored in the memory. Usually they store all the training
data in the memory during the learning phase, and defer all the
essential computation until the prediction phase. Examples of such
techniques are $k$-nearest neighbor, kernel regression and locally
weighted regression. If setting $k$ to $n$ (the number of data
points) and optimizing weights by gradient descent, $k$-nearest
neighbor turns into kernel regression, while locally weighted
regression generalized kernel regression, not just obtains local
average values. In general, irrelevant features are often killers
for instance-based approaches. But ANNs can be trained directly on
problems with hundreds or thousands of inputs. Instance-based
learning methods can fit low dimensional, very complex functions
very accurately while ANNs require considerable tweaking to do
this. When adding new data, training is almost free for
instance-based learning methods, but ANNs and SVMs need retraining
the data.

We put forward a kernel regression method to estimate photometric
redshifts. This paper is organized as follows. In Section~2 we
describe the data we use. A brief overview of kernel regression is
addressed in Section 3. Section 4 illustrates the results and
discussion, and the conclusion is presented in Section 5.

\section{DATA}
The Sloan Digital Sky Survey (SDSS, York et al. 2000) is the most
ambitious astronomical survey ever undertaken. When completed, it
will provide detailed optical images covering more than a quarter of
the sky, and a 3-dimensional map of about a million galaxies and
quasars, with a dedicated 2.5-meter telescope located on Apache
Point, New Mexico. The first stage of SDSS is already complete (with
DR5). It has imaged 8,000 square degrees in five bandpasses ($u, g,
r, i, z$) and measured spectra of more than 675,000 galaxies, 90,000
quasars and 185,000 stars. In its second stage, SDSS will carry out
three new surveys in different research areas, such as the nature of
the universe, the origin of galaxies and quasars and the formation
and evolution of the Milky Way. In order to construct a
representative sample set, we collected all objects satisfying the
follow criteria from SDSS Data Release 5 (Adelman-McCarthy et al.
2007). All following mentioned magnitudes are magnitudes corrected
by Galaxy extinction using the dust maps of Schlegel et al. 1998.
After these restrictions that the spectroscopic redshift confidence
must be greater than or equal to 0.95, and the redshift flags should
be zero, we obtained a sample containing 399,929 galaxies.

The photometry properties discussed below are available in all
five SDSS bandpasses ($ugriz$), however the $r$-bandpass values
for these quantities are usually applied for the $r$-band result
generally has the lowest error and gives more consistent results
(Way \& Srivastava 2006). The Petrosian 50\% (90\%) radius is the
radius where 50\% (90\%) of the flux of the object contributes.
$r50$ is Petrosian 50\% radius in $r$ band, $r90$ is Petrosian
90\% radius in $r$ band. The ratio of these quantities is called
Petrosian concentration index $c$=$r90/r50$, which is an indicator
of the galaxy type: early-type galaxy with $c>$ 2.5 and late-type
galaxy with $c<$ 2.5 (Strateva et al. 2001). The Petrosian Radii
are also utilized together with a measure of the profile type from
the SDSS photometric pipeline reduction named fracDeV. fracDeV
results from a linear combination of the best exponential and de
Vaucouleus profiles that are fit to the image in each band.
fracDeV is a floating point number between zero and 1. fracDeV is
closely related to galaxy type while it is 1 for a pure de
Vaucouleurs profile typical of early-type galaxies and zero for a
pure exponential profile typical of late-type galaxies. eClass is
a spectroscopic parameter giving the spectral type from a
principal component analysis, which is a continuous value ranging
from about -0.5 (early-type galaxies) to 1 (late-type galaxies).

\section{KERNEL REGRESSION}
\subsection{Overview of the algorithm}

Kernel regression (Watson, 1964; Nadaraya, 1964) belongs to the
family of instance-based learning algorithms, which simply store
some or all of the training examples and ``delay learning" till
prediction time. Given a query point $\mathbf{x_q}$, a prediction is
obtained using the training samples that are ``most similar" to
$\mathbf{x_q}$. Similarity is measured by means of a distance metric
defined in the hyper-space of $V$ predictor variables. Kernel
regressors obtain the prediction for a query point $\mathbf{x_q}$,
by a weighted average of the $y$ values of its neighbors. The weight
of each neighbor is calculated by a function of its distance to
$\mathbf{x_q}$ (called the kernel function). These kernel functions
give more weight to neighbors that are nearer to $\mathbf{x_q}$. The
notion of neighborhood (or bandwidth) is defined in terms of
distance from $\mathbf{x_q}$. The prediction for query point
$\mathbf{x_q}$ is obtained by
\begin{equation}
y_q={\frac{\sum\limits_{i=1}^{N}K({\frac{D(\mathbf{x_i},\mathbf{x_q})}{h})}\times
y_i}{{\sum\limits_{i=1}^{N}K({\frac{D(\mathbf{x_i},\mathbf{x_q})}{h}})}}}
\end{equation}
where $D(.)$ is the distance function between two instances; $K(.)$
is a kernel function; $h$ is a bandwidth value; $(\mathbf{x_i},
y_i)$ are training samples; $\mathbf{x_i}$ and $\mathbf{x_q}$ are
vectors; $N$ is the number of datapoints used in the model. In this
paper, we use Euclidian distance and Gaussian kernel function.
$\mathbf{x_i}$ is the feature for each training sample, $y_i$ is the
spectroscopic redshift for each training set sample, $y_q$ is the
redshift of each query sample.

\subsection{Bandwidth determination}

One important design decision when using kernel regression is the
choice of the bandwidth $h$. The larger $h$ results in the flatter
weight function curve, which indicates that many points of
training set contribute quite evenly to the regression. As the $h$
tends to infinity the predictions approach the global average of
all points in the database. If the $h$ is very small, only closely
neighboring datapoints make a significant contribution. If the
data is relatively noisy, we expect to obtain smaller prediction
errors with a relatively larger $h$. If the data is noise free,
then a small $h$ will avoid smearing away fine details in the
function. There exists mature algorithms for choosing the
bandwidth for kernel regression that minimize a statistical
measure of the difference between the true underlying distribution
and the estimated distribution. Usually bandwidth selection in
regression is done by cross-validation (CV) or the penalized
residual sum of squares.

Cross-validation is the statistical method of dividing a sample of
data into subsets such that the analysis is initially performed on a
single subset, while the other subset(s) are retained for subsequent
use in confirming and validating the initial analysis. $M$-fold
cross-validation is one important cross-validation method. The data
is divided into $M$ subsets of (approximately) equal size. Each
time, one of the $M$ subsets is used as the test set and the other
$M-$1 subsets are put together to form a training set.
Cross-validation is designed to choose the bandwidth by minimizing
the cross-validation score CV($h$) defined by
\begin{displaymath}
CV(h)=\frac{1}{M}[\frac{1}{k_{1}}\sum_{i=0}^{k_{1}}(y_{1i}-\widehat{y}_{1i})^2+\frac{1}{k_{2}}\sum_{i=0}^{k_{2}}(y_{2i}-\widehat{y}_{2i})^2
\end{displaymath}
\begin{equation}
+...+\frac{1}{k_{M}}\sum_{i=0}^{k_{M}}(y_{Mi}-\widehat{y}_{Mi})^2]
\end{equation}
where $y_{ji}$ is the spectroscopic redshift for each test set
sample, $\widehat{y}_{ji}$ is the predicted photometric redshift of
each test sample, $k_{j}$ is the number of objects in each subset
($j=1,2,...,M$), $M$ is the number of subsets for cross-validation.
In general, the $k_{j}$ values are identical. Here we adopt 10-fold
cross-validation for the bandwidth choice, i.e. $M$=10, firstly
divide the sample of 399,929 galaxies into 10 subsets, then 9
subsets of 10 subsets are taken as training set and the rest subset
as testing set for ten times.

We adopt the sample described in Section 2, applying four color
indexes ($u-g$, $g-r$, $r-i$ and $i-z$) and spectroscopic redshifts
as input parameters. Then we implement kernel regression on this
sample and compute the 10-fold cross-validated score for different
bandwidths in Table 1. As shown by Table~1, the cross-validated
score CV($h$) reaches the minimum $5.559\times10^{-4}$ when $h$ is
equal to 0.02. Therefore, 0.02 has been assigned to the optimal
fixed bandwidth for the sample in this case.

\subsection{Input pattern selection}

In this work, we choose the input parameters using the Akaike
Information Criterion (AIC). AIC (Akaike 1974) is a measure of the
goodness of fit of an estimated statistical model. The AIC
methodology attempts to find the model that best explains the data
with a minimum of free parameters. In the general case, AIC is
\begin{equation}
AIC=-2\ln\emph{L$_{\rm max}$}+2k
\end{equation}
where \emph{L$_{\rm max}$} is the maximized likelihood function,
and \emph{k} is the number of free parameters in the model.

The purpose of model selection is to identify a model that best fits
the available data set. A model is better than another model if it
has a smaller AIC value. When a model approach the lowest values of
AIC, the model is regarded as the best model. Several recent works
in astrophysics have used AIC for model selection (e.g. Liddle 2004,
2007). In Section 4.1, AIC will be used to select the optimal input
pattern.

\begin{table}
\begin{minipage}{1.\linewidth}
\caption{Bandwidth determination using the cross-validated (CV)
method }
 \label{Tab:table-1}
\begin{center} \begin{tabular}{lc}
\hline \hline $h$ & CV($h$)($\times 10^{-4})$ \\
\hline
0.010 & 5.668 \\
0.015 & 5.574   \\
\textbf{0.020} & \textbf{5.559} \\
0.025 & 5.620 \\
0.030 & 5.725 \\
0.035 & 5.831 \\
0.040 & 5.973 \\
0.045 & 6.112 \\
0.050 & 6.264 \\
0.055 & 6.426 \\
0.060 & 6.601 \\
0.065 & 6.794 \\
0.070 & 6.990 \\
0.075 & 7.195 \\
0.080 & 7.410 \\
0.085 & 7.638 \\
0.090 & 7.877 \\
 \hline \hline
\end{tabular}
\end{center}
\vskip 0.3cm
\begin{tabular}{l}

\end{tabular}
\end{minipage}
\end{table}

\section{RESULTS AND DISCUSSION}
\subsection{RESULTS}

One advantage of the empirical training set approach to photometric
redshift estimation is that additional parameters can be easily
incorporated. More parameters (e.g. $r50$, $r90$, fracDeV etc.) may
be taken as inputs. In order to study which parameters influence the
accuracy of predicting photometric redshifts, we probe different
input patterns to estimate photometric redshifts. According to the
bandwidth choice criterion described in Section 3.2, we compute the
10-fold cross-validation scores and get the optimal bandwidth values
corresponding to different situations, as shown in Table 2. In order
to determine which input pattern is best, we use the AIC criterion
to investigate this problem.
\begin{table}
\caption{rms errors, optimal bandwidths and AIC for different
input parameters}
 \label{Tab:table-2}
\begin{center}
\begin{tabular}{lccc}
\hline \hline
Input Parameters$^{*}$ & $\sigma_{\rm rms}$ & $h$ & AIC \\
\hline
$ugriz$           &  0.0215    & 0.025 & 64.259\\
$ugriz+r50+r90$   &  0.0247    & 0.070 & 84.282\\
$ugriz+$fracDeV$\_r$ & 0.0223  & 0.035 & 69.242\\
$ugriz$+eclass    &  0.0198    & 0.025 & 54.548\\
color             &  0.0220    & 0.020 & 67.558\\
color$+r$         &  0.0206    & 0.030 & 58.933\\
color$+r+c$       &  0.0206    & 0.035 & 58.656\\
color$+r+r50+r90$ &  0.0226    & 0.050 & 70.206\\
color+fracDeV$\_r$&  0.0220    & 0.025 & 67.149\\
color+$ugriz$     &  0.0210    & 0.040 & 60.961\\
color+eclass      &  0.0189    & 0.025 & 49.503\\
\hline \hline
\end{tabular}
\end{center}
NOTE.----$r50$ is Petrosian 50\% radius in $r$ band, $r90$ is
Petrosian 90\% radius in $r$ band, fracDeV$\_r$ is fracDeV in $r$
band, color is the color indexes, i.e. $u-g$, $g-r$, $r-i$, $i-z$,
and $c=r90/r50$.
\end{table}

When implementing kernel regression to predict photometric
redshifts, 260,000 galaxies are randomly regarded as training set
and the rest are as test set. The rms deviations, optimal bandwidth
and AIC for different input patterns are listed in Table 2. Table 2
shows that rms error is different for each input pattern while the
corresponding optimal bandwidth and AIC are different, too.
Nevertheless AIC has the same trend as rms error, i.e. AIC increases
with the increase of rms error and decreases with the decline of rms
error. When AIC approaches minimum, the input pattern is considered
as the best input pattern, vice versa. As a result, the best input
pattern is four colors ($u-g$, $g-r$, $r-i$, $i-z$) and eClass when
rms error amounts to 0.0189. The next better input pattern is five
magnitudes and eClass when rms error is 0.0198. Then the good input
pattern is four colors and $r$ magnitude when the rms scatter is
0.0206. The result with only five magnitudes is better than that
with only four colors but worse than that with four colors and $r$
magnitude. For five magnitudes as inputs, the performance of kernel
regression decreases when adding $r50$ and $r90$ or fracDeV$\_r$
except eClass. Similarly, for four colors or four colors and $r$
magnitude as inputs, the performance becomes worse when also
considering $r50$ and $r90$ or fracDeV$\_r$. The performance adding
the Petrosian concentration index $c$ hasn't improved compared with
only four colors and $r$ magnitude as inputs. The result with four
colors and five magnitudes is superior to that only with colors or
only with magnitudes, however it is worse than that with four color
and $r$ magnitude. Therefore when applying kernel regression to
predict photometric redshifts, we find the parameters except
magnitudes and color indexes, such as $r50$, $r90$, fracDeV$\_r$ and
$c$, contribute little information, however eClass is important and
effective.

Figure 1 shows the comparison of the known spectroscopic redshift
with the calculated photometric redshift from the test data using
kernel regression with the input pattern of color+eClass.
Considering color+$r$ as the inputs, the fractions of predicted
photometric redshifts exceeding $\pm 3\sigma$ and $\pm 4\sigma$
error bar with the loss of estimation are 2.10\% and 1.03\%,
respectively. With color+eClass as the inputs, the fractions
including the loss occupy 2.11\% and 1.28\%, separately. The loss of
estimation refers to the points whose photometric redshifts can not
be measured due to their distance to neighbors beyond the optimal
window width of kernel regression.

\begin{figure}
\includegraphics[width=8cm,height=8cm,clip]{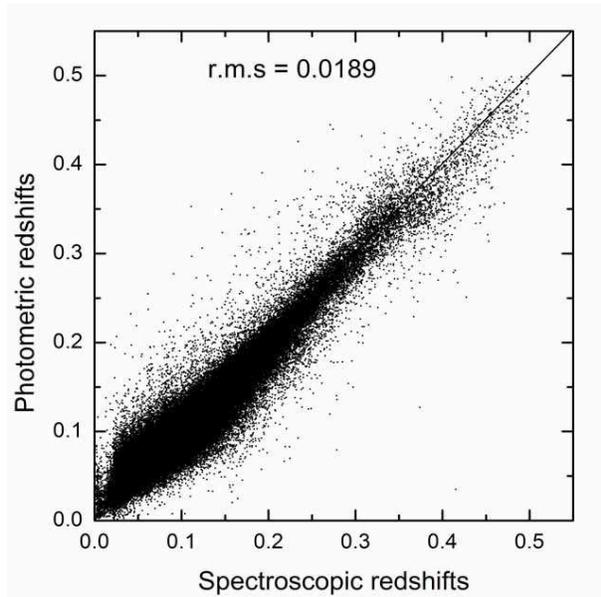}
\caption[fig4] {Comparison between spectroscopic and photometric
redshifts. 260,000 galaxies are regarded as training set. 139,929
galaxies are as test set (plotted). The input parameters are
$u-g$, $g-r$, $r-i$, $i-z$ and eClass.}
\end{figure}

Although eClass is not strictly photometric, it is applicable to use
this parameter to estimate photometric redshifts when galaxies have
low S/N spectra, or they have weak absorbtion or emission lines.
Moreover it is helpful for the statistical study of a large galaxy
sample without detailed spectra information. In addition, eClass may
be estimated with color indexes or magnitudes, just like following.
The parameter eClass is a continuous parameter ranging from
approximately -0.5 (early type galaxies) to 1 (late type galaxies),
indicating spectral type in the SDSS spectroscopic catalog. We use
the same sample to estimate eClass rather than redshifts with kernel
regression. Based on the result as listed in Table 2, we choose the
best input pattern of color+$r$ except the patterns with eClass. The
rms scatter is $\sigma_{\rm rms}$ = 0.0337, as shown in Figure 3.
Other researchers have done similar works, for example, Wadadekar
(2005) utilized support vector machines (SVMs) to predict the
photometric eClass using 10,000 objects from SDSS Data Release 2 and
the rms scatter of eClass estimation $\sigma_{\rm rms}$ = 0.057;
Collister $\&$ Lahav (2004) obtained $\sigma_{\rm rms}$ = 0.052 by
artificial neural networks (ANNs) for the eClass estimation with
64,175 objects from SDSS Data Release 1.

\begin{figure}
\includegraphics[width=8cm,height=8cm,clip]{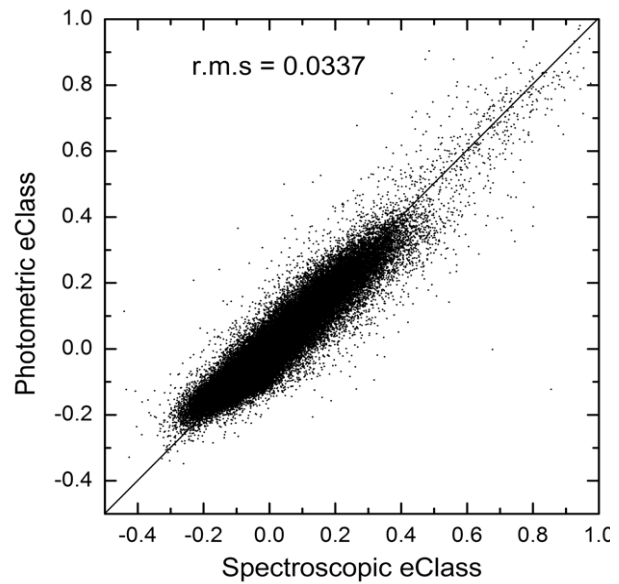}
 \caption[fig5]
 {Spectroscopic eClass vs. calculated photometric eClass for 139,929 galaxies from the SDSS DR5 with kernel
 regression. The input parameters are
$u-g$, $g-r$, $r-i$, $i-z$ and $r$.
  } \label{iz}
\end{figure}

From Table 2, we can draw a conclusion that spectral type is an
important parameter for determining photometric redshifts. In
order to further study how the spectral type influences the
accuracy of measuring photometric redshifts, the sample is divided
into two parts according to the criterion that early-type galaxy
is $c>$ 2.5 and late-type galaxy is $c<$ 2.5 (Strateva et al.
2001). Thus 251,794 early-type galaxies and 148,135 late-type
galaxies are obtained in our sample. Then we implement kernel
regression on the two sets separately. When taking $u-g$, $g-r$,
$r-i$, $i-z$ as inputs and $h$=0.02, the rms dispersion of
photometric redshifts is $\sigma_{\rm rms}$=0.0197 for early-type
galaxies and $\sigma_{\rm rms}$=0.0247 for late-type galaxies, the
rms scatter ($\sigma_{rms}^{mix}$) for the mixed sample adds up to
0.0215. The computation of $\sigma_{rms}^{mix}$ refers to Equation
(4).
\begin{equation}
\sigma_{rms}^{mix}=\sqrt{\frac{1}{N_{1}+N_{2}}(\sum_{i=1}^{N1}(y^{E}_{i}-\widehat{y}_{i}^{E})^2+
\sum_{i=1}^{N2}(y_{i}^{L}-\widehat{y}_{i}^{L})^2)}
\end{equation}
where $y^{E}_{i}$ and  $y^{L}_{i}$ are the spectroscopic redshift
for early-type and late-type galaxies, respectively;
$\widehat{y}_{i}^{E}$ and $\widehat{y}_{i}^{L}$ are the predicted
photometric redshift of early-type and late-type galaxies,
separately. $N_{1}$ is the number of early-type galaxies; $N_{2}$
is the number of late-type galaxies.

When taking $u-g$, $g-r$, $r-i$, $i-z$ and $r$ as inputs and
$h$=0.03, the rms error of photometric redshifts is $\sigma_{\rm
rms}$=0.0186 for early-type galaxies and $\sigma_{\rm rms}$=0.0230
for late-type galaxies, the mixed rms error is 0.0204. Considering
four color indexes and eClass as inputs and $h$=0.025, the rms
scatter is $\sigma_{\rm rms}$=0.0164 for early-type galaxies and
$\sigma_{\rm rms}$=0.0222 for late-type galaxies, the mixed rms
error amounts to 0.0187. The rms scatter with two parts of sample
outperforms that without separating the sample, as shown in Table~3.
For early-type galaxies, the rms deviation of photometric redshift
measurement is very satisfactory. Table 3 further indicates that the
parameter of eClass related to spectral type is robust and
significant to determine the photometric redshifts and it is also
helpful to improve the accuracy of photometric redshifts with the
separation of galaxies into early-type ones and late-type ones. In
addition, AIC values approach minimum simultaneously with
color+eClass as the inputs for early-type and late-type galaxies.
Therefore, in our case, color+eClass is the best input pattern to
determine photometric redshifts while color+$r$ is the next better
one.

\newsavebox{\tablebox}
\begin{table}
\caption{Comparison of the accuracy for the separated sample with
that for the original sample}
 \label{Tab:table-3}
\begin{lrbox}{\tablebox}
\begin{tabular}{lccccccc}
\hline \hline
Input Parameters & $\sigma_{\rm rms}^{E}$ & AIC$^{E}$ & $\sigma_{\rm rms}^{L}$ & AIC$^{L}$  &  $\sigma_{\rm rms}^{mix}$ & $\sigma_{\rm rms}$ \\
\hline
color           &  0.0197 & 38.52  & 0.0247 & 35.79 & 0.0215 & 0.0220 \\
color$+r$       &  0.0186 & 36.96  & 0.0230 & 33.86 & 0.0204 & 0.0206\\
color+eClass    &  0.0164 & 30.33  & 0.0222 & 31.94 & 0.0187 & 0.0189\\
\hline \hline
\end{tabular}
\end{lrbox}
\scalebox{0.8}{\usebox{\tablebox}}
 NOTE.----$\sigma_{\rm rms}^{E}$ is $\sigma_{\rm rms}$ for
early-type galaxies; $\sigma_{\rm rms}^{L}$ is $\sigma_{\rm rms}$
for late-type galaxies; $\sigma_{\rm rms}^{mix}$ for the whole
sample; AIC$^{E}$ and AIC$^{L}$ are AIC values for early-type and
late-type galaxies, respectively. $\sigma_{\rm rms}$ is taken from
Table 2.
\end{table}

\subsection{DISCUSSION}

At present there have been many works on the algorithms to
determining photometric redshifts. Each method has its pros and
cons. For ANNs, we need to make a decision about the optimal
network architecture. More complex network architectures we have
more accurate result. ANNs allow a closer fit to the data, but are
subject to the danger of overfitting. In addition, adding layers
or nodes to the network, training time will increase remarkably
(Wadadekar 2005). Comparing to ANNs, SVMs simplifies the training
process, only need to choose the kernel function rather than the
architecture. Even simple Gaussian function can give a good
performance. However, the adjustments of lots of parameters
require prior knowledge. Correlation between parameters makes the
regulating process more complicated. Although linear or non-linear
polynomial regression is easy to communicate with astronomers, the
systematic deviation is large (Brunner et al. 1997; Wang, Bahcall
\& Turner 1998; Budav$\acute{a}$ri et al. 2005; Hsieh et al. 2005;
Connolly et al. 1995). In recent years, a combination of HyperZ
with the Bayesian marginalization was proposed by Benitez (2000).
The dispersion of photometric redshifts using this combination
technique was significantly improved. The results using Bayesian
technique have been ameliorated, nevertheless, the application of
this method can introduce unrealistic effects in some studies.
Therefore, this approach can be an alternative option when one is
dealing with no spectral data.

\begin{table}
\caption{Various photometric redshift approaches and accuracies}
 \label{Tab:table-4}
\begin{center}
\begin{tabular}{lccccc}
\hline \hline
Method Name & $\sigma_{\rm rms}$ & Data set & Input parameters \\
\hline
CWW$^{1}$         &  0.0666    & SDSS-EDR & $ugriz$ \\
Bruzual-Charlot$^{1}$    &  0.0552    & SDSS-EDR & $ugriz$  \\
Interpolated$^{1}$&0.0451& SDSS-EDR & $ugriz$\\
Polynomial$^{1}$    &  0.0318    & SDSS-EDR & $ugriz$ \\
Kd-tree$^{1}$      &  0.0254 & SDSS-EDR & $ugriz$  \\
ClassX$^{2}$  &  0.0340    & SDSS-DR2 & $ugriz$  \\
SVMs$^{3}$    &  0.027 & SDSS-DR2 & $ugriz$   \\
  & 0.0230 & SDSS-DR2 & $ugriz+r50+r90$ \\
ANNs$^{4}$       &  0.0229   & SDSS-DR1 & $ugriz$ \\
Polynomial$^{5}$     & 0.025 & SDSS-DR1,GALEX & $ugriz+nuv$ \\
Kernel Regression & 0.0215 & SDSS-DR5 & $ugriz$  \\
   & 0.0206 & SDSS-DR5 & color+$r$  \\
   & 0.0189 & SDSS-DR5 & color+eclass  \\
 \hline \hline
\end{tabular}
\end{center}
NOTE.---- SDSS-EDR = Early Data Release (Stoughton et al. 2002),
SDSS-DR1 = Data Release 1 (Abazajian et al. 2003), SDSS-DR2 = Data
Release 2 (Abazajian et al. 2004), SDSS-DR5 = Data Release 5
(Adelman-McCarthy et al. 2007). $r50$ is Petrosian 50\% radius in
$r$ band, $r90$ is Petrosian 90\% radius in $r$ band, fracDeV$\_r$
is fracDeV in $r$ band, color is the color indexes, i.e. $u-g$,
$g-r$, $r-i$, $i-z$. \\(1) Csabai et al. 2003; (2) Suchkov,
Hanisch \& Margonet 2005;\\(3) Wadadekar 2005; (4) Collister \&
Lahav 2004; (5) Budav$\acute{a}$ri et al. 2005.
\end{table}

With large and deep photometric surveys are carried out, it seems
that kernel regression will offer some significant advantages over
other approaches, as shown in Table 4. The performance of kernel
regression to predict photometric redshifts is comparable to ANNs
and SVMs, superior to Kd-tree, ClassX and polynomial regression,
and more preferable than CWW and Bruzual-Charlot (Wadadekar 2005;
Collister \& Lahav, 2004; Csabai et al. 2003; see their Tables 1).
A major problem for empirical training-set method is the
difficulty in extrapolating to regions where the input parameters
are not well represented by the training data. But for kernel
regression, even though a few high-redshift galaxies exists in the
sample, one can appropriately adjust bandwidth to obtain much more
accurate redshifts. In addition, compared to other training-set
methods, kernel regression has another advantage that it needn't
retraining when a new query point appears.

\section{CONCLUSION}

We have presented an instance-based learning method called kernel
regression to predict photometric redshifts of galaxies with the
data from SDSS broadband photometry. Important work in kernel
regression is how to determine the bandwidth. We use 10-fold
cross-validation to choose the optimal bandwidth. Our experiments
show that the optimal bandwidth is different for different input
parameters, the color+eClass pattern is the best when the rms
error of photometric redshift estimation adds up to 0.0189, the
$ugriz$+eClass is better when the rms error is 0.0198. Except
these two situations, the color$+r$ pattern is the best when the
rms scatter is 0.0206. The parameters, such as $r50$, $r90$,
fracDeV$\_r$ and $c$, contribute little information, however
eClass shows much importance. Moreover kernel regression achieves
high accuracy to predict photometric redshifts for early-type
galaxies and the photometric eClass. For ANNs, the more parameters
considered, the accuracy of photometric redshifts is higher (Way
\& Srivastava 2006; Li et al. 2006). While for kernel regression
and SVMs, the accuracy is satisfactory only when appropriate
parameters are chosen. To our satisfaction, kernel regression is
able to measure photometric redshifts of galaxies, accurately.
This is helpful to construct the sample of galaxies for the study
of cosmology with minimal contamination from objects at seriously
incorrect redshifts. Similarly kernel regression may be applied to
predict photometric redshifts of quasars.

Kernel regression has a number of flexibilities. It is possible to
make different queries with not only different kernel widths $h$,
but also different distance metrics, with subsets of attributes
ignored, or with some other distance metrics such as Manhattan
distance, Canberra distance. It is also possible to apply the same
technique with different kernel functions for classification
instead of regression. Unlike the traditional training methods,
its best merit is the ability to make predictions with different
parameters without needing a retraining phase, moreover it doesn't
seriously depend on the size of sample. Nevertheless it has the
obvious disadvantage of instance-based learning that is a
significant computational cost on large data sets. In the future
work we will explore different functions or other kinds of
distance metric for kernel regression on the regression problems.
In addition, we may use multiresolution instance-based learning as
suggested by Deng \& Moore (1995). This method succeeds in
reducing the cost of instance-based learning, moreover it has two
advantages: flexibility to work throughout the local/global data;
the ability to make predictions with different parameters without
needing a retraining phase.

 \vskip 0.5cm

\section{acknowledgements}

The authors are very grateful to the anonymous referee whose
insightful and detailed comments led to the improvement of our work.
This paper is funded by National Natural Science Foundation of China
under grant No.90412016, No.60603057 and 10778623.

Funding for the SDSS and SDSS-II has been provided by the Alfred P.
Sloan Foundation, the Participating Institutions, the National
Science Foundation, the U.S. Department of Energy, the National
Aeronautics and Space Administration, the Japanese Monbukagakusho,
the Max Planck Society, and the Higher Education Funding Council for
England. The SDSS Web Site is http://www.sdss.org/.

The SDSS is managed by the Astrophysical Research Consortium for the
Participating Institutions. The Participating Institutions are the
American Museum of Natural History, Astrophysical Institute Potsdam,
University of Basel, University of Cambridge, Case Western Reserve
University, University of Chicago, Drexel University, Fermilab, the
Institute for Advanced Study, the Japan Participation Group, Johns
Hopkins University, the Joint Institute for Nuclear Astrophysics,
the Kavli Institute for Particle Astrophysics and Cosmology, the
Korean Scientist Group, the Chinese Academy of Sciences (LAMOST),
Los Alamos National Laboratory, the Max-Planck-Institute for
Astronomy (MPIA), the Max-Planck-Institute for Astrophysics (MPA),
New Mexico State University, Ohio State University, University of
Pittsburgh, University of Portsmouth, Princeton University, the
United States Naval Observatory, and the University of Washington.


\end{document}